\documentclass[aps,preprint]{revtex4}%
\usepackage{amsfonts}
\usepackage{amsmath}
\usepackage{amssymb}
\usepackage{graphicx}%
\providecommand{\U}[1]{\protect \rule{.1in}{.1in}}

\begin{document}
\preprint{ }
\title[Short title for running header]{The Spectral Phase-Amplitude Representation of a Wave Function.}
\author{George Rawitscher}
\affiliation{Physics Department, University of Connecticut, Storrs, CT 06269}
\pacs{PACS number}

\begin{abstract}
The phase and amplitude (Ph-A) of a wave function vary slowly and
monotonically with distance, in contrast to the wave function that can be
highly oscillatory. Hence an attractive feature of the Ph-A representation is
that it requires far fewer meshpoints than for the wave function itself. In
1930 Milne developed an equation for the phase and the amplitude functions (W.
E. Milne, Phys. Rev. \textbf{35}, 863 (1930)), and in 1962 Seaton andPeach (M.
J. Seaton and G. Peach, Proc. Phys. Soc. 79 1296 (1962)) developed an
iterative method for solving Milne's Ph-A equations. Since the zero'th order
term of the iteration is identical to the WKB\ approximation, there is a close
relationship between the Ph-A and the WKB representations of a wave function.
The objective of the present study is to show that a spectral Chebyshev
expansion method to solve Seaton and Peach's iteration scheme is feasible, and
requires very few meshpoints for the whole radial interval. Hence this method
provides an economical and accurate way to calculate wave functions out to
large distances. In a numerical example for which the potential decreased
slowly with distance as $1/r^{3}$, the whole radial range of $[0-2000]$
covered with $301$ mesh points (and Chbyshev basis functions). The first order
iteration of the Ph-A wave function was found to have an accuracy better than
$1\%$, and was always more accurate than the WKB wave function.

\end{abstract}
\startpage{1}
\endpage{102}
\maketitle

\section{Introduction}

When the Phase-Amplitude (Ph-A) method was first introduced by Milne in 1930
\cite{MILNE} , and then taken up by many authors, see Ref. \cite{KORSCH}, the
main motivation was the paucity of numerical mesh points required, compared to
the calculation of the wave function itself. This is because both phase and
amplitude functions are monotonic and slowly varying, as opposed to the wave
function itself that can be highly oscillatory. This point was verified by
many authors, in particular by Calogero and Ravenhall \cite{RAVEN} who state
that the solution for the phase is more stable than the solution of the wave
function. An additional argument in favor of the (Ph-A) representation is that
it lends itself to analytic expressions to address particular problems. For
example, the Ph-A representation facilitates the incorporation of the effect
of long range potentials \cite{ROBICH}, \cite{DEHMER}\ or the calculation of
resonances \cite{KORSCH}. It is also helpful in the quantum defect calculation
of atomic wave functions \cite{GREENE}, the calculation of Gaunt Factors
\cite{WIM}, as well as the description of an electron with an ion embedded in
a plasma \cite{RITCHIE}, among others. The Ph-A description of a carrier wave
in radio or television also plays a significant r\^{o}le in the
compactification of the signal transmission in the field of Information
Technology. An additional advantage of the Ph-A representation is that it
provides a method to improve the WKB approximation of a wave function, an
important point since the WKB approximation \cite{WKB} has led, over the
years, to a much improved understanding of the solution of the Schr\"{o}dinger Eq.

The Ph-A representation consists in writing a wave function $\psi(r)$ in the
form
\begin{equation}
\psi(r)=y(r)\sin[\phi(r)], \label{y sin}%
\end{equation}
where $y$ is the amplitude and $\phi$ is the phase, and $r$ the distance from
the origin. If an overlap matrix element
\begin{equation}
M=\int_{0}^{\infty}\psi_{1}(r)U(r)\psi_{2}(r)dr. \label{M12}%
\end{equation}
between two wave functions is required, then in the finite difference method
of obtaining integrals, both $\psi_{1}$ and $\psi_{2}$ have to be calculated
on a sufficiently fine mesh, which can be time consuming and prone to errors.
However, the Ph-A representation can provide an estimate of $M$ by decomposing
the integrand of the overlap matrix element into a slowly oscillating (S)
\ and a fast oscillating (F) part%
\begin{equation}
M=M^{(S)}-M^{(F)}. \label{MFMS}%
\end{equation}
The decomposition makes use of a trigonometric identity for the product of two
sine functions with the result%
\begin{equation}
M^{(F,S)}=\frac{1}{2}\int_{0}^{\infty}y_{1}(r)U(r)y_{2}(r)[\cos(\phi_{1}%
\pm \phi_{2})]dr. \label{IF IS}%
\end{equation}
The matrix element $M^{(S)}$ can be calculated on a small set of radial mesh
points since the integrand oscillates slowly. Further, since $M^{(F)}%
<M^{(F)},$ a rough estimate for $M$ is provided by $M^{(S)}$ alone. Here
$U(r)$ is an overlap function that depends on the physics application envisaged.

In 1962 Seaton and Peach \cite{Seaton} presented an iterative scheme to solve
Milne's non-linear differential equation \cite{MILNE} for the amplitude and
phase. It is the purpose of the present work to implement this iterative
method by means of a spectral \cite{SPECTRAL} expansion of the amplitude in
terms of Chebyshev polynomials. A further purpose is to examine the accuracy
of the resulting Ph-A wave function by comparison with the direct solution of
the Schr\"{o}dinger equation for the wave function, the latter also obtained
by an accurate spectral integral equation method \cite{CHEB}, denoted as $IEM$
in what follows. The combination of both objectives have not been presented
previously. The great advantage of a spectral expansion is that the
calculations utilize all the support points located in a given partition
simultaneously, with the result that the errors are shared uniformly across
the partition in the case of Chebyshev expansions \cite{BOYD}. For the present
numerical examples the calculation is done in one great radial partition,
extending from $r=0$ to $r=2000$, containing $201$ Chebyshev support points.
By contrast, other algorithms (such as finite elements, finite differences, or
the $IEM$ method described below) have to divide such a large radial interval
into a number of partitions, with the result that the error from one partition
is propagated into the adjoining one, the last partition having the largest
error \cite{POWER}. In addition, for calculations that require\ the storage of
many wave functions with high precision \cite{DEREV}\ the use of the Ph-A
representation can be very advantageous because the amount of storage required
can be substantially smaller than what is needed for other algorithms.

In section II the iterative method is explained, section III contains details
of the computational spectral method, section IV presents the results,
including error estimates and suggestions for improvements, and finally the
Summary and Conclusions are presented in section V.

\section{Iterative solution of Milne's Phase-Amplitude equation.}

Milne \cite{MILNE} and others have derived a non linear equation for the
amplitude $y$ and phase $\phi$ \ for a partial wave functions $\psi$, which is%
\begin{equation}
d^{2}y/dr^{2}+k^{2}y=V_{T}\ y+\frac{k^{2}}{y^{3}} \label{NL_SCHR}%
\end{equation}
where the total potential $V_{T}$\ is%
\begin{equation}
V_{T}=L(L+1)/r^{2}+V(r). \label{Vext}%
\end{equation}
Here $V(r)$ is the atomic or nuclear potential (including the Coulomb
potential), $L$ is the orbital angular momentum quantum number, and the
nonlinearity is given by the last term in Eq. (\ref{NL_SCHR}). The phase
$\phi(r)$ is obtained from the amplitude $y(r)$ according to \cite{MILNE}
\begin{equation}
\phi(r)=\phi(r_{0})+k\int_{r_{0}}^{r}[y(r^{\prime})]^{-2}\ dr^{\prime},
\label{phase}%
\end{equation}
but it can also be obtained without the knowledge of $y$ \cite{WIM}. The Eq.
(\ref{NL_SCHR}) has been solved non-iteratively in the past by using some form
of a finite difference computational method, such as one of Milne's
predictor-corrector methods \cite{AS}, or \cite{RITCHIE} by a Bulirsch-Stoer
limit method \cite{Burl-Stoer}, none of which will be used in the present study.

The iterative method of Seaton and Peach \cite{Seaton} consists in rewriting
Eq. (\ref{NL_SCHR}) in the form%
\begin{equation}
\frac{k^{2}}{y^{4}}=w+\frac{1}{y}\frac{d^{2}y}{dr^{2}} \label{Seaton1}%
\end{equation}
where
\begin{equation}
w(r)=k^{2}-V_{T}, \label{w}%
\end{equation}
and calculating the solution of Eq. (\ref{Seaton1}) by means of the iteration
\cite{Seaton}
\begin{equation}
\frac{k}{y_{n+1}^{2}}=[w+\frac{1}{y_{n}}\frac{d^{2}y_{n}}{dr^{2}}%
]^{1/2},\  \ n=0,1,2,... \label{Siter}%
\end{equation}
Here $n$ denotes the order of the iteration, and the initial value of $y$ is
given by the WKB approximation \cite{WKB}%
\begin{equation}
\frac{k}{y_{0}^{2}}=w^{1/2}. \label{WKB}%
\end{equation}
The advantage of formulating the iteration according to Eq. (\ref{Siter}) is
that $y$ varies slowly and monotonically with $r$ for large distances$,$ and
hence $(1/y_{n})d^{2}y_{n}/dr^{2}$ is small compared to $w$. Near the origin
of $r$ this term may become large, but a numerical solution of Eq.
(\ref{Siter}) still converges very well according to Ref. \cite{Seaton}. At
large distances where $w(r)\rightarrow k^{2}$ the amplitude $y$ automatically
approaches unity. The Eq. (\ref{phase}) combined with the first order result
(\ref{WKB}) is equivalent to the WKB approximation, and hence the iteration
scheme (\ref{Siter}) provides a method to iteratively improve the WKB approximation.

\section{Computational method}

The spectral computational method consist in expanding the function $y$ into a
series of $N+1$ Chebyshev polynomials $T_{s}(x),$ $s=0,1,2,..N$ ,
\begin{equation}
y(x)=\sum_{s=0}^{N}a_{s}T_{s}(x). \label{aT}%
\end{equation}
That expansion is inserted into Eq. (\ref{Siter}), and the corresponding
coefficients $a_{s}$are obtained by solving a matrix equation \cite{CHEB},
\cite{CiSE}. The driving term of this equation\ is the known right hand side
of Eq. (\ref{Siter}), which is also expanded in terms of Chebyshev
polynomials. Since the Chebyshev polynomials are defined in the interval
$-1\leq x\leq1$ the quantities defined in the radial interval $0\leq r\leq
r_{\max}$ are mapped into the $x-$variable by a linear transformation.
According to the spectral methods the $x$- mesh points are the $N+1$ zeros of
$T_{N+1}$. The expansion cutoff value $N$ is set arbitrarily, but once chosen,
the location and number of support points on the $x-$axis is determined.
Extensive use is made of the Clenshaw-Curtis matrix method (CC) \cite{CC} that
relates the values of a function evaluated at the $N+1$ mesh-points to the
expansion coefficients $a_{s}$ of that function, and vice-versa, by a simple
known matrix \cite{CHEB} relation.

\begin{figure}
[ptb]
\begin{center}
\includegraphics[
height=2.4163in,
width=3.2145in
]%
{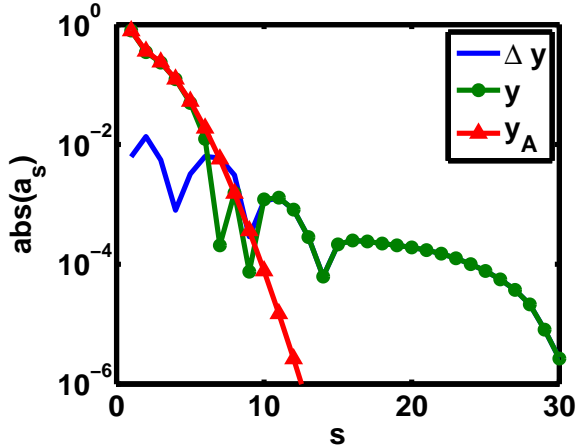}%
\caption{The absolute values of Chebyshev expansion coefficients as a function
of the Chebyshev index $s$ for $y_{WKB},\Delta y$, and $y_{A}.$}%
\label{absay}%
\end{center}
\end{figure}

The second order derivatives of $y_{n}$ are obtained by replacing the $T_{s}$
in Eq. (\ref{aT}) by their respective second derivatives, and keeping the
coefficients $a_{s}$ unchanged.
\begin{equation}
\frac{d^{2}y}{dr^{2}}=\sum_{s=0}^{N}a_{s}\frac{d^{2}T_{s}(x)}{dx^{2}}%
(\frac{dx}{dr})^{2} \label{D2y}%
\end{equation}
By using the expression $T_{s}(x)=\cos(s\theta)$, $s=0,1,..N$ , in terms of
$\theta$, where $x=\cos(\theta)$, one obtains after some trigonometric
transformations
\begin{equation}
\frac{d^{2}T_{s}(x)}{dx^{2}}=\frac{s}{\sin^{2}(\theta)}\left[  \frac
{\sin((s-1)\theta)}{\sin(\theta)}-(s-1)\cos(s\theta)\right]  ,\text{
\  \  \ }s=0,1,2,.. \label{D2T}%
\end{equation}
In order to obtain these derivatives in $r-$space, it is sufficient to use
$dx/dr=2/(b_{2}-b_{1})$, where $b_{2}$ and $b_{1}$ are the right and left
extrema of the radial interval. However the calculation of the second order
derivative in Eq. (\ref{Siter}) introduces errors \cite{CiSE}, and these
errors increase as $N$ is made larger. This feature is the major source of
error in the present procedure, since the derivatives of Chebyshev polynomials
increase\ substantially with the order $s$ of the polynomial, and may overcome
the decrease with $s$ of the coefficients $a_{s}$. For example, for $s=16$ and
$x=-1,$ $d^{2}T_{s}(x)/dx^{2}=2\times10^{4}.$ For this reason, a balance
between the desired accuracy that increases with $N$, and the error in the
second order derivative of $T_{s}$ has to be achieved. In order to overcome
the difficulty described above, the function $y$ is approximated by an
analytical function $y_{A}$, plus a remainder function $\Delta y.$
\begin{equation}
y_{n}=y_{A}+\Delta y_{n} \label{deltay}%
\end{equation}
with
\begin{equation}
y_{A}(r)=1-\exp[(r-r_{S})/\alpha] \label{ANAL}%
\end{equation}
The second order derivative of $y_{A}$ is obtained analytically, and the
second order derivative of $\Delta y$ is obtained by using Eq. (\ref{D2T}).
The decrease of the expansion coefficients of $\Delta y$ relative to $y$ is
illustrated in Fig. \ref{absay}\ for the case that $y$ has the WKB value, as
discussed below.\  \ The figure shows that the values for the expansion of
$\Delta y$ are smaller by two orders of magnitude than the coefficients for
$y$ for small values of the index $s$, and remain small. This feature permits
one to evaluate the second order derivative of $\Delta y$ by using Eq.
(\ref{D2y}) without undue loss of accuracy, while the same would not have been
the case for the second order derivative of $y.$ The values of the parameters
$r_{S}$ \ and $a$ in Eq. (\ref{ANAL}) are listed in Table \ref{TABLE1}
\begin{table}[tbp] \centering
\begin{tabular}
[c]{|l||l|l|}\hline
$\  \ k$ & $\ r_{S}$ & $\  \alpha$\\ \hline \hline
$0.005$ & $500$ & $5$\\ \hline
$0.01$ & $250$ & $5$\\ \hline
$0.1$ & $60$ & $10$\\ \hline
\end{tabular}
\caption{The values of the parameters in Eq. (\ref{ANAL})}\label{TABLE1}%
\end{table}%
The integral in Eq. (\ref{phase}) required to calculate the phase $\phi$ is
performed by a Gauss-Chebyshev method \cite{CHEB}, \cite{CiSE} that is well
suited to this type of spectral expansion since it only requires the values of
the expansion coefficients $a_{s}$. Situations that involve imaginary local
wave numbers and the respective turning points, as is the case in the presence
of repulsive barriers, are postponed to a future study.

The calculations are done with MATLAB on a desk PC using an Intel TM2 Quad,
with a CPU Q 9950, a frequency of 2.83 GHz, and a RAM\ of 8 GB. The
calculation uses typically $N+1=31$ Chebyshev polynomials for the calculation
of $u$. The computing time for the iterative spectral part of the calculation,
compared with the IEM calculation, both carried out in the whole radial
interval $[0,2500]$ is given in Table \ref{TABLE2}. The computing time for the
Ph-A iterations depends only on the number of Chebyshev functions $N+1$,
regardless of the size of the radial interval, and depends weakly on the value
of $k.$ \ For $N=200$, and performing one iteration, the calculation requires
approximately $0.18\ s$. That does not include the time to interpolate the
results to a fine equidistant radial mesh. Interpolating $y$ and $\phi$ to an
equi-spaced radial mesh size of step length $h=0.1$ depends on the size of the
radial interval. For the radial interval $[0,40]$ the fine mesh interpolation
requires $0.8\ s,$ and for the radial interval $[40,2000]$ the interpolation
takes $170\ s$ to $180\ s$. However, the calculation of the slowly oscillating
part $M^{(S)}$ of an overlap matrix element (\ref{IF IS}) can be done by using
the Gauss-Chebyshev integration method \cite{CiSE}, which does not require the
interpolation to an equi-spaced radial mesh, and is expected to take
approximately $0.30\ s$ for obtaining both of the two wave functions and also
$M^{(S)}.$
\begin{table}[tbp] \centering
\begin{tabular}
[c]{|l||l|l|}\hline
$k$ & Ph-A (s) & $\ IEM(s)$\\ \hline \hline
$0.01$ & $0.18$ & $0.20$\\ \hline
$0.1$ & $0.18$ & $0.29$\\ \hline
\end{tabular}
\caption{Computation times, as explained in the text}\label{TABLE2}%
\end{table}%
.

\section{Results}

The feasibility of the present approach will be demonstrated by means of an
example, for which the potential $V_{T}$ is everywhere attractive and has a
long range tail proportional to $r^{-3}$. Three wave numbers are used,
$k=0.1,\ 0.01,$ and $0.005$, the radial region extends from $r=0$ to $r=2000$,
and the orbital angular momentum is $L=0.$ In Eq. (\ref{NL_SCHR}) the factor
$\hslash^{2}/2m$ has already been divided into the potential and into the
energy $k^{2}$, so that both are given in units of inverse length squared. The
unit of distance can be either $fm$ for nuclear physics applications, or the
Bohr radius $a_{0}$ for atomic physics applications, but will not be
explicitly indicated.

The potential is the sum of a Woods-Saxon form, Eq. (\ref{C6}), to which is
added a $r^{-3}$ tail, whose singularity at the origin is smoothly removed by
an analytic mapping procedure, Eqs. (\ref{C7}-\ref{C8})
\begin{figure}
[ptb]
\begin{center}
\includegraphics[
height=2.0211in,
width=2.687in
]%
{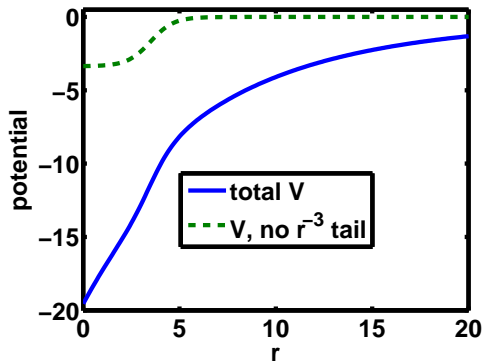}%
\caption{(Color online) The solid line illustrates the potential used for the
numerical examples. The units are in inverse length squared, since the
potential, in energy units, has been multiplied by the factor $2m/\hbar^{2}.$
The dashed line indicates the Woods-Saxon potential to which is smoothly added
a $1/r^{3}$ "tail", as described in the text.}%
\label{FIGpot}%
\end{center}
\end{figure}
\begin{equation}
V_{WS}(r)=-3.36\ /\  \left[  1+\exp \{(r-3.5)/0.6\} \right]  \label{C6}%
\end{equation}%
\begin{equation}
V_{3}(r)=-1.6224\times10^{4}/\mathcal{R}^{3} \label{C7}%
\end{equation}%
\begin{equation}
\mathcal{R(}r\mathcal{)}=r/[1-\exp(-r/10)] \label{C8}%
\end{equation}%
\begin{equation}
V=V_{WS}+V_{3}. \label{C9}%
\end{equation}
The values of these potentials are appropriate for atomic physics applications
\cite{BA}. The reason this $1/r^{3}$ long range nature was chosen, is because
this case did not get addressed successfully by means of a Born-approximation
method \cite{BA}, while it is well described in the present study. The Woods
Saxon part and the total potential $V$ are illustrated respectively by the
dashed and solid lines in Fig. \ref{FIGpot}. The long-ranged nature of this
potential is such that at $r=2500$ the value of $V$ is $\simeq10^{-6}.$ The
corresponding wave function is highly oscillatory at small distances, with an
amplitude that varies substantially with distance, as is illustrated in Fig.
\ref{FIG18}.
\begin{figure}
[ptb]
\begin{center}
\includegraphics[
height=1.9216in,
width=2.5547in
]%
{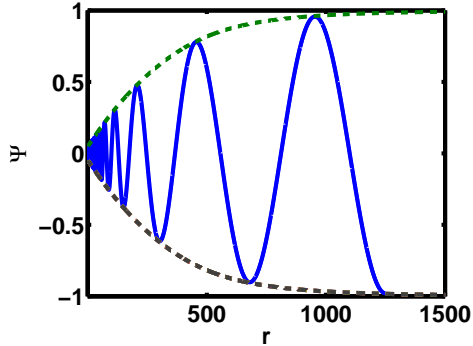}%
\caption{(Color online). The agreement of the amplitude $y,$ shown by the
dashed lines, with the extrema of the solid line representing the \ $IEM$ wave
function, for $k=0.01$ inverse length.}%
\label{FIG18}%
\end{center}
\end{figure}
The corresponding amplitude $y(r)$ is illustrated by the dashed lines in Fig.
\ref{FIG18}$.$ It is in good agreement with the wave function calculated by
the spectral IEM method \cite{CHEB}, denoted as $IEM$, and shown in Fig.
\ref{FIG18} by the solid line. Noteworthy is the fact that only $201$
expansion terms in Eq. (\ref{aT}) have been used to calculate the amplitude
for the whole radial interval $[0,2000]$. The phase functions $\phi(r)$, based
on Eq. (\ref{phase}), are illustrated in Fig. \ref{FIGxx} for two values of
the wave number $k$. It is not clear wether the phase function obtained here
is identical to the one examined by Calogero in his excellent book
\cite{CALOG}, because the equations each one obeys are very different from
each other, although asymptotically they must agree.
\begin{figure}
[ptb]
\begin{center}
\includegraphics[
height=1.8844in,
width=2.5062in
]%
{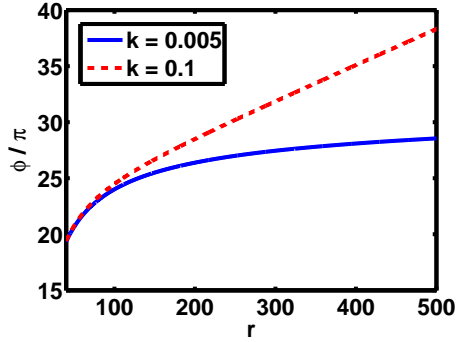}%
\caption{(Color online). The phase functions divided by $\pi$ for the
potential shown in Fig. \ref{FIGpot} for two different values of the wave
number $k.$ For the larger value of $k$ the wave function has more
oscillations, hence the phase function increases more rapidly with $r.$}%
\label{FIGxx}%
\end{center}
\end{figure}
Unless otherwise noted, the numerical results described further below are
carried out only to the first iteration order $n=1$, since the main pupose of
the study was to establish the feasibility of the method. Additional
iterations could proceed along the lines of Eq. (\ref{Siter}), but a more
effective method could be established by subtracting the WKB amplitude from
$y_{n}$, i.e., $z_{n}=y_{n}-y_{WKB},$ and since $z_{n}<<y_{WKB}$ the resulting
equation for $z_{n}$ could be linearized$.$An example of the good agreement
between the IEM and the Ph-A wave functions is illustrated in Fig
\ref{FIG20}.
\begin{figure}
[ptb]
\begin{center}
\includegraphics[
height=2.1923in,
width=2.9162in
]%
{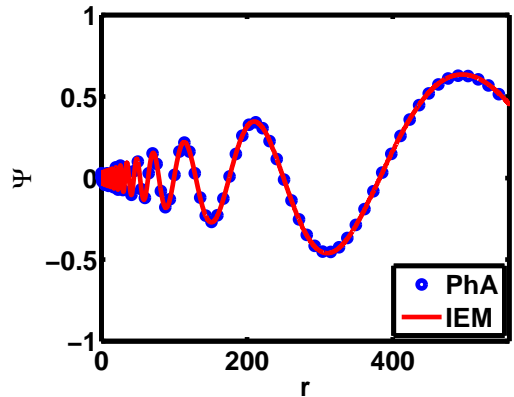}%
\caption{(Color on line). The solid line is the IEM wave function, while the
open circles illustrate the Ph-A wave function results at the Chebyshev
support points with $N=300$, for $k=0.005$. The Ph-A calculation extends ftom
$r=0$ to $r=2000$, but only the radial interval $[0,500]$ is shown.}%
\label{FIG20}%
\end{center}
\end{figure}

\begin{figure}
[ptb]
\begin{center}
\includegraphics[
height=2.0418in,
width=2.7146in
]%
{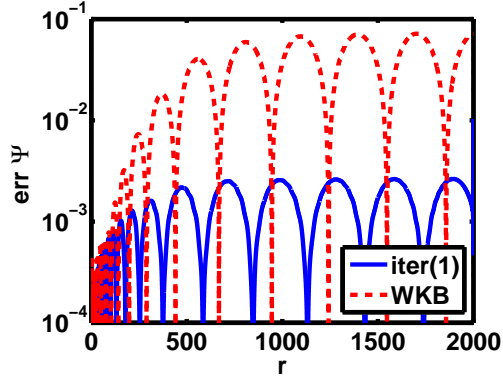}%
\caption{(Color on line). The error of the Ph-A wave function for $k=0.01$,
using $201$ chebyshev expansion functions for the whole radial interval $0\leq
r\leq2000.$}%
\label{FIG22}%
\end{center}
\end{figure}

\begin{figure}
[ptb]
\begin{center}
\includegraphics[
height=1.9052in,
width=2.5339in
]%
{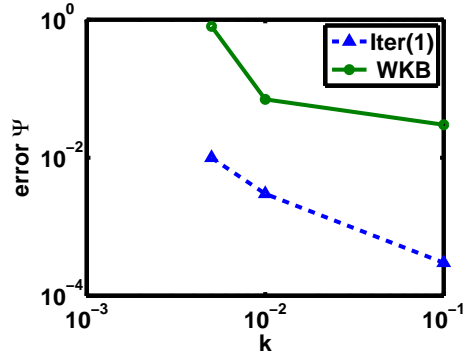}%
\caption{(Color on line). The error of the Ph-A wave functions, obtained by
comparison with the $IEM$ wave functions, for three values of the wave number
$k,$ (in units of inverse length) for large distances in the vicinity of
$r=2000$ . For the small distances, in the vicinity of $r=20$, all errors are
of the same magnitude, less than $10^{-3}.$ }%
\label{FIGerror}%
\end{center}
\end{figure}

An evaluation of the error of the wave function is obtained by plotting the
absolute value of the difference of the Ph-A and the IEM wave functions. The
result for the case $k=0.01$ is illustrated in Fig. \ref{FIG22}, which shows
that the agreement between the Ph-A and IEM wave functions for the large
distances is close to $0.1\  \%,$ while the error of the WKB wave function is
larger than $1\%$ . For the smaller distances, $0<r<40,$ both the WKB and the
Ph-A wave functions have an error less than $10^{-3}$.
The values of the errors for the WKB and Ph-A wave functions for the three
values at the large distances are summarized in Fig. \ref{FIGerror}%
The general conclusion for this particular numerical case studied is that in
the smaller radial intervals the WKB approximation is slightly less accurate
than the Ph-A method for the smaller distances, but is less accurate by more
than an order of magnitude for the large radial distances. This latter result
shows the value of the present form of the Ph-A method, which provides
\ further corrections to the WKB results, requiring very few mesh-points

\subsection{Overlap Integrals}

\begin{figure}
[ptb]
\begin{center}
\includegraphics[
height=1.6181in,
width=2.1517in
]%
{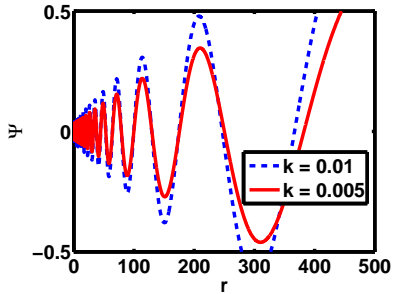}%
\caption{(Color on line). The two wave functions used in the calculation of
the matrix $M$, defined in Eq. (\ref{M12}). Both have unit amplitude at
$r=2500$.}%
\label{FIG26}%
\end{center}
\end{figure}

\begin{figure}
[ptb]
\begin{center}
\includegraphics[
height=1.5921in,
width=2.1171in
]%
{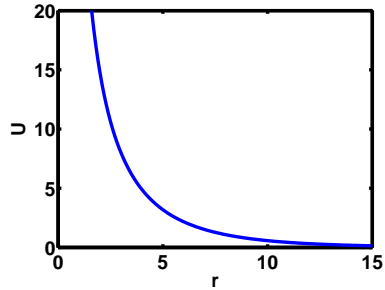}%
\caption{(Color online). The overlap function $U$ that occurs in the matrix
element $M$, defined in Eq. (\ref{M12}). Because of the factor $1/r$, it
becomes $\infty$ at $r=0.$ The units of $U$ are $(1/length)^{2}$}%
\label{FIG27}%
\end{center}
\end{figure}

\begin{figure}
[ptb]
\begin{center}
\includegraphics[
height=1.8066in,
width=2.4016in
]%
{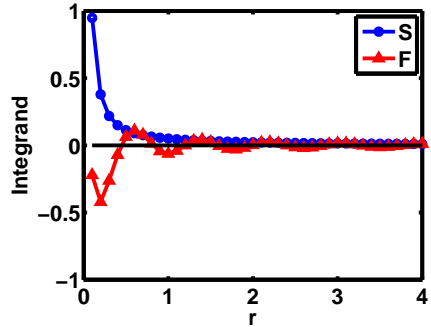}%
\caption{(Color online). The integrands of the matrix elements $M^{(F)}$ and
$M^{(S)}$. Due to the oscillation of the integrand of $M^{(F)}$, it is clear
that $M^{(F)}<M^{(S)}.$ }%
\label{FIG28}%
\end{center}
\end{figure}

An example of the calculation of matrix elements by means of the Ph-A method
will be presented below. The two wave functions $\psi_{1}$ and $\psi_{2}$ are
solutions of the one-dimensional radial Schr\"{o}dinger equation with the
potential $V$ defined in Eqs. (\ref{C6}) to (\ref{C9}), for different wave
numbers $\ k=0.01$ and $0.005$, respectively (in units of inverse length). The
two wave functions have different amplitudes, but nearly the same phases at
distances where $|V|\ >\ k^{2},$ as illustrated in Fig. \ref{FIG26}.
The overlap potential $U$ is taken from Eq. (4) of Ref. \cite{RITCHIE}, and
represents the screened interaction of an electron with an ion embedded in a
plasma. It is composed of a sum of exponentials divided by the radial distance
$r$, and is illustrated in Fig.\ref{FIG27}.
It has a $1/r$ singularity at $r\rightarrow0.$ Using the Ph-A representation
of $\psi_{1}$ and $\psi_{2}$, the integrand of the overlap integral separates
into a fast oscillating and slowly oscillating parts, Eqs. (\ref{IF IS}), as
described above. These integrands are illustrated in Fig. \ref{FIG28}.
The approximate values of $M^{(F)}$and $M^{(S)}$ are $-0.073$ and $0.258$. As
expected, the integrand of $M^{(F)}$ is more oscillatory than the integrand of
$M^{(S)},$ and hence $|M^{(F)}|\ <|M^{(S)}|.$ Hence a crude estimate of $M$ is
given by $M^{(s)},$ which can be calculated directly within the Ph-A
representations, without the necessity to interpolate to small radial meshes.

\section{ Summary and conclusions}

This is the first time that the iterative method of Seaton and Peach
\cite{Seaton} was successfully combined with a spectral Chebyshev expansion of
the amplitude $y$ in solving the non linear equation of Milnes \cite{MILNE}
for the amplitude representation of a wave function. The difficulty with the
Chebyshev expansion of $y$ in obtaining the second order derivative of $y$ was
overcome by the simple procedure of decomposing $y$ into an analytic part
$y_{A}$ plus a remainder $\Delta y.$ The second order derivative of $y_{A}$ is
obtained analytically, and since $\Delta y<<y,$ the second order derivative of
$\Delta y,$ given by its Chebyshev expansion, caused no difficulty. For a
numerical example that contains a long range potential tail proportional to
$r^{-3}$, it was found that $300$ basis functions sufficed to span the entire
radial domain from the origin to $r=2000,$ and the resulting Ph-A wave
function was accurate to $0.1\%$ in the whole domain. An interesting feature
of the Seaton and Peach's iteration scheme is that the zero'th order
approximation is identical to the WKB approximation. The accuracy of the
latter was in some of the cases less than $1\%,$ but the first iteration
increased the accuracy to $0.1\%$, as illustrated in Fig. \ref{FIGerror}.

The Ph-A method is expected to be very useful for a) the calculation of
overlap matrix elements that involve highly oscillatory wave functions, b) to
obtain the long range value of wave functions in cases where the conventional
solutions of the Schr\"{o}dinger equation may be inadequate, and c) to provide
a very economical method to store wave functions. The present results open the
way to generalize the Ph-A method to scattering cases where barriers are
present, to bound states, or to the situation of coupled channel equations for
which only the final phases in each channel are required.

The author is indebted to Dr. Ionel Simbotin for calling attention to the Ph-A
representation, and for stimulating conversations.


\begin{thebibliography}{99}                                                                                               %


\bibitem {MILNE}W. E. Milne, Phys. Rev. \textbf{35}, 863 (1930);

\bibitem {KORSCH}M. J. Korsch and H. Laurent. J. Phys. B:\ At. Mol. Phys
\textbf{14, }4213 (1981);

\bibitem {RAVEN}F. Calogero and D. G. Ravenhall, Nuovo Cimento \textbf{32},
1755 (1964);

\bibitem {ROBICH}F. Robicheaux, U. Fano, M. Cavagnero, and D. A. Harmin, Phys.
Rev. A, \textbf{35}, 3619 (1987);

\bibitem {DEHMER}J. L. Dehmer and U. Fano, Phys. Rev. \textbf{A 2}, 304 (1970);

\bibitem {GREENE}C. H. Greene, A. R. P. Rau and U. Fano, Phys. Rev. \textbf{A
26}, 2441 (1982);

\bibitem {WIM}B. Wilson, C. Iglesias, and Mau Chen, J. Quant. Spectrosc.
Radiat. Transf. \textbf{81}, 499 (2003);

\bibitem {RITCHIE}A. B. Ritchie and A.K. Bhatia, Phys. Rev. E \textbf{69},
035402(R), 2004;

\bibitem {WKB}H. Jeffreys an B. S. Jeffreys, \emph{Methods of Mathematical
Physics, }Cambridge University Press, NY (1966); H. A. Kramers, Z. Physik,
\textbf{39}, 828 (1926);

\bibitem {Seaton}M. J. Seaton and G. Peach Proc. Phys. Soc. 79 1296 (1962) doi:10.1088/0370-1328/79/6/127;

\bibitem {SPECTRAL}A. Deloff, Ann. Phys. (NY) \textbf{322}, 1373--1419
(2007);L. N. Trefethen, \emph{Spectral Methods in MATLAB}, (SIAM,
Philadelphia, PA, 2000);

\bibitem {CHEB}R. A. Gonzales, J. Eisert, I Koltracht, M. Neumann and G.
Rawitscher, J. of Comput. Phys. \textbf{134}, 134 (1997); R. A. Gonzales,
S.-Y. Kang, I. Koltracht and G. Rawitscher, J. of Comput. Phys. \textbf{153},
160-202 (1999);

\bibitem {BOYD}Y. L. Luke, \emph{Mathematical Functions and their
Approximations,}\textbf{ (}Academic Press, NY, 1975); John P. Boyd,
\emph{Chebyshev and Fourier Spectral Methods,} 2nd revised ed. (Dover
Publications, Mineola, NY, 2001);

\bibitem {POWER}J. Power and G. Rawitscher, Phys. Rev. E\textbf{ 86},066707 (2012);

\bibitem {DEREV}V. A. Dzuba, A. Derevianko, J. of Phys. B: Atomic, Molecular
and Optical Physics, \textbf{43}, 074011 (2010); V. A. Dzuba, A. Derevianko,
V. V. Flambaum, Phys. Rev. \textbf{A 86}, 054501 (2012); S. G. Porsev, A.
Derevianko, Phys. Rev. \textbf{A 74}, 020502, ( 2006); M. S. Safronova, S. G.
Porsev, U. I. Safronova, M. G. Kozlov, C. W. Clark, Phys. Rev. \textbf{A 87},
012509 (2013);

\bibitem {AS}M. Abramowitz and I. Stegun, eds., (Handbook of Mathematical
Functions, Dover, 1972), Eq. 25.5.13;

\bibitem {Burl-Stoer}W. H. Press, S. A. Teukolsky, W. T. Vetterling, B. P.
Flannery, (2007). "Section 17.3. Richardson Extrapolation and the
Bulirsch-Stoer Method". Numerical Recipes: \emph{The Art of Scientific
Computing (3rd ed.)}. New York: Cambridge University Press. ISBN 978-0-521-88068-8;

\bibitem {CiSE}G. Rawitscher and I. Koltracht, Computing Sci. Eng. \textbf{7},
58 (2005);

\bibitem {CC}C.C. Clenshaw, and A.R. Curtis, Numer. Math., 1960,\emph{
}\textbf{2, }197;

\bibitem {BA}G. Rawitscher, Phys. Rev. \textbf{A} \textbf{87}, 032708 (2013);

\bibitem {CALOG}F. Calogero, \emph{Variable Phase Approach to Potential
Scattering, }Academic Press (NY, 1967)
\end{thebibliography}
\end{document}